\documentclass[%
 reprint,
 amsmath,amssymb,
 aps, prl
 ]{revtex4-2}
\usepackage{graphicx}
\usepackage{dcolumn}
\usepackage{float}
\usepackage{svg}
\usepackage{mathalfa}
\usepackage{braket}
\usepackage{siunitx}
\usepackage{bm}
\DeclareMathOperator\erf{erf}

\begin{document}


\title{Pushing the limits of negative group velocity}

\author{A. J. Renders} \author{D. Gustavsson} \author{M. Lindén}  \author{A. Walther} \author{S. Kröll} \author{A. Kinos} \author{L. Rippe}
 \affiliation{%
  Department of Atomic Physics, Lund University.\\
 }%

\date{\today}

\begin{abstract}
Distortion free negative group velocity pulse propagation is demonstrated in a rare-earth-ion-doped-crystal (RE) through the creation of a carefully designed spectral absorption structure in the inhomogeneous profile of Eu:YSO and subsequently inverting it. The properties of the RE system make it particularly well suited for this since it supports the creation of very sharp, arbitrarily tailored spectral features, which can be coherently inverted by a single pulse thanks to the long coherence time of the transition. All together these properties allow for a large time advancement of pulses without causing distortion. A pulse advancement of 13$\%$ with respect to the pulse full-width-half-maximum was achieved corresponding to a time-bandwidth product of 0.06. This to our knowledge is the largest time-bandwidth product achieved, with negligible shape distortion and attenuation. Our results show that the rare-earth platform is a powerful test bed for superluminal propagation in particular and for dispersion profile programming in general. 

\end{abstract}

\maketitle
\begin{figure*}
    \centering
    \includegraphics[width = \textwidth]{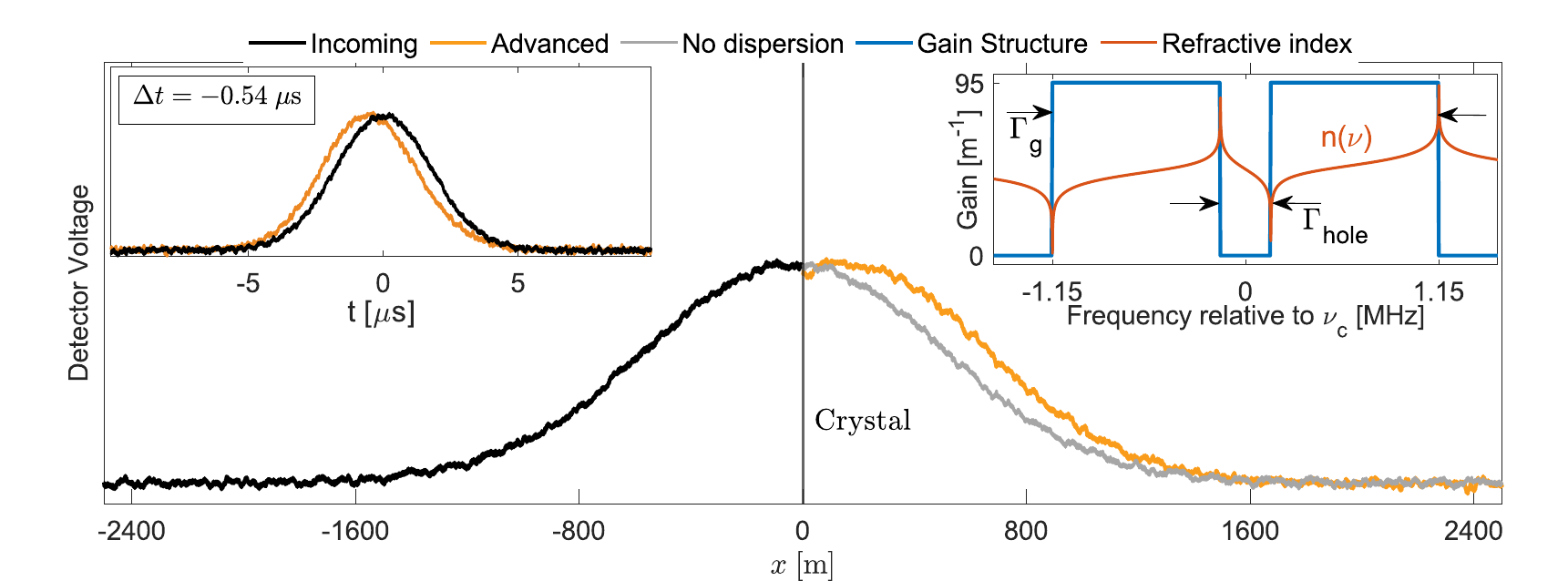}
    \caption{The spatial propagation of the pulse as it approaches the crystal. For clarity, the incoming and no dispersion pulses are scaled to the same intensity as the probe that propagates through the crystal. Unnormalised data can be found in the supplement. The position of the crystal is indicated by the vertical line. Here it can be seen that the advanced pulse peak comes out of the material before the incoming pulse peak enters. In the left inset the pulses are shown in time, with the advancement $\Delta t$. The $t_{\text{FWHM}}$ ratio between the $t_{\text{FWHM}}$ for the advanced and reference pulse was 0.984. The pulse has a $t_{\text{FWHM}}$ of 4.2~\unit{\mu s}. The right inset shows the gain structure used in the experiment and its effect on the refractive index. The gain structure width, $\Gamma_{\text{g}}$~=~2.3~\unit{MHz}, and hole width, $\Gamma_{\text{hole}}$~=~300~\unit{kHz} are indicated.}
    \label{fig:fastlight_result}
\end{figure*}

Manipulation of the group velocity, $v_g$, of pulses has been a subject of study for several decades. This manipulation is done by controlling the dispersion of the medium in which the pulse propagates. Controlling dispersion is vital in various applications, e.g. chirped pulse amplification \cite{Strickland1985} and in optical fibres \cite{Nielsen2005}. Reducing $v_g$ has led to an interesting field of research with a number of  applications such as laser stabilisation~\cite{Zhao2009,Horvath2022} and medical imaging \cite{Bengtsson2019}. When working at the other end of the spectrum of $v_g$, in particular increasing $v_g$ beyond the vacuum light speed $c$, or when $v_g$ becomes negative, things become more abstract and less intuitive. One cannot violate causality, but negative group velocities can be achieved and can be described by the Kramers-Kronig relations~\cite{Fang2017}. A number of experiments, starting as early as the 1960s, have been conducted to show that one can go into a regime where $v_g > c$ and can even go negative i.e. the pulse peak exits the material earlier than it entered~\cite{Basov1966, Chu1982, Segard1985, Chiao1993, Chiao1997, Rajan2015, Zhou2013}. The methods used include using the dispersion in an absorbing medium near a transparency \cite{Rajan2015} and non-linear amplification \cite{Basov1966}. However, all these early experiments showed either a great deal of distortion of the pulses that were advanced in time~\cite{Basov1966, Chu1982} or the pulses were heavily attenuated \cite{Segard1985, Chiao1993, Chiao1997, Rajan2015, Zhou2013}. This left room for different interpretations of the result and the physical meaning of what was seen. In the strong attenuation case, for example, it was argued that the pulse was not advanced, but rather selectively attenuated such that the pulse peak appears to be shifted forward, but is still completely contained within the original pulse envelope.

To address the concerns raised regarding the physical interpretation, an undistorted, advanced pulse needed to be produced. It was suggested by Steinberg et al.~\cite{Steinberg94} that one could achieve undistorted pulse propagation through a transparent optical window between two gain regions. This was subsequently shown experimentally by Wang et al.~\cite{Wang2000}, using a Raman pumped system in a caesium vapour cell in $\Lambda$ configuration. However, the initial publication showed the advanced pulse still had some distortion. This among other factors made Ringenmacher and Mead \cite{Ringenmacher2000} question the interpretation. To them, the 3$\%$ shape error with only 62~\unit{ns}, 1.6$\%$, advancement on a pulse with $t_{\text{FWHM}}=$~3.7~\unit{\mu s}, and hence a time-bandwidth product of 0.007, seemed insufficient for a proper conclusion. Later Wang et al. published a second paper where they presented a pulse with similar advancement but without the previously observed distortion~\cite{Dogariu2001}.\newline

In this work, we demonstrate almost an order of magnitude improvement in terms of time-bandwidth product, and relative pulse advancement. A Gaussian probe pulse with time full width half maximum (FWHM), $t_{\text{FWHM}}$~=~4.2~\unit{\mu s}, was used in these experiments. This pulse was advanced by 0.543~\unit{\mu s} which corresponds to 13$\%$ of the $t_{\text{FWHM}}$, see FIG~\ref{fig:fastlight_result}, left inset. The relation of the FWHM of a Gaussian pulse in time and frequency is $t_{\text{FWHM}} = 0.44/ \nu_{\text{FWHM}}$, hence $\nu_{\text{FWHM}}$~=~105~\unit{kHz}. This gives a time-bandwidth product, $|\Delta t|\cdot\nu_{\text{FWHM}}$, of 0.06, the corresponding group refractive index $n_g$ is hence -6.6$\cdot10^{3}$. The shape was well maintained, the $t_{\text{FWHM}}$ stayed within 1.6$\%$ and the pulse power within 3.0$\%$ (upper bound estimation) of a reference pulse.

We realised these results through the creation of two sharp, closely spaced gain regions in the inhomogeneous profile of a Europium-doped YSO crystal. This leads to a region of strong and linear anomalous dispersion between the gain regions, see FIG~\ref{fig:fastlight_result}, right inset. The rare-earth system allows for much sharper gain features than were created in the caesium vapour cell \cite{Wang2000}, and a group refractive index that is strongly negative. We are not aware of any previous result of a pulse time advancement with this large a time-bandwidth product, with as little distortion or attenuation. 

Light pulses travel at the group velocity $v_g$ \cite{Saleh}: 
\begin{equation}
    v_g(\nu)   = \frac{c}{n(\nu) + \nu \frac{d n(\nu)}{d\nu}}
\end{equation}
With the speed of light in vacuum $c$, the refractive index $n(\nu)$ and the optical frequency $\nu$ \cite{Siegman}. From this expression, one can see that the refractive index and the change of the refractive index with frequency, the dispersion, influence the group velocity. 
In close vicinity of resonances of an ion, i.e. close to regions with absorption or gain, very strong dispersion can be achieved, see FIG \ref{fig:fastlight_result}, right inset.  This behaviour can be described using the susceptibility, the imaginary and real parts of which govern the absorption and refractive index response respectively. Using this, one can calculate the dispersion.

The real part of the susceptibility can be computed fairly simply from the ion distribution generating the gain, which has a width of $\Gamma_{g}$, and the hole width, $\Gamma_{\text{hole}}$. The index of refraction as a function of frequency $n(\nu)$ was derived for this ion distribution. Since the region of interest is the transparent region around the centre frequency of the hole (where the fast light pulse propagates), a Taylor expansion is applied to get a linear approximation of $n(\nu)$: 
\begin{equation}
    n(\nu) = n_{\text{host}} + \frac{\alpha c}{\pi^2\nu_c}\left(\frac{1}{\Gamma_{\text{hole}}}- \frac{1}{\Gamma_{g}}\right)(\nu-\nu_c)
    \label{eq:nu}
\end{equation}
With the background refractive index, $n_{\text{host}}$, and the central optical frequency, $\nu_c$. Here contributions to the refractive index stemming form beyond the frequency range in the right inset in FIG~(\ref{fig:fastlight_result}) are ignored. It should be noted that in the case of an inverted population the absorption coefficient, $\alpha$, will change sign and become a gain coefficient. A full derivation can be found in the supplement. The group refractive index can then be found by inserting equation~(\ref{eq:nu}) into the equation for the group refractive index $n_g = n(\nu) + \nu\frac{dn(\nu)}{d\nu}|_{\nu=\nu_{c}}$ which is evaluated at the centre frequency. When the dispersion is large, $n_{\text{host}}$ is negligible and the dispersion term completely dominates the denominator, hence  the group velocity $v_{g}(\nu=\nu_c)$:
\begin{equation}
    v_g = \frac{\pi^2}{\alpha}\left(\frac{1}{\Gamma_{\text{hole}}}-\frac{1}{\Gamma_g}\right)^{-1}
\end{equation}

\begin{figure*}
    \centering
    \includegraphics[width = \textwidth]{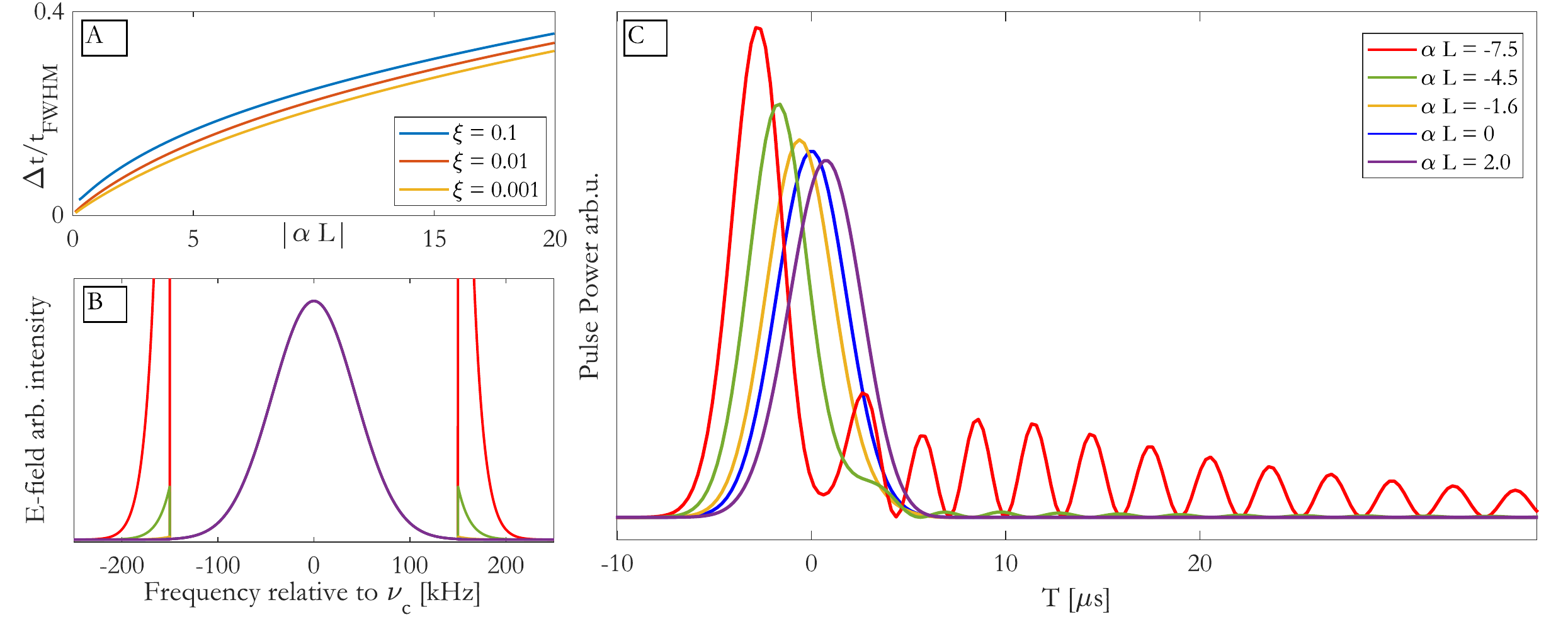}
    \caption{(a) Relative pulse advancement as a function of $\alpha L$ at fixed $\Gamma_{\text{hole}}$ (=300~\unit{kHz}) for different degrees of tolerance in the pulse power increase. (b) Increase in pulse area of the electric field caused by the gain structure for different values of $\alpha L$ corresponding to those in (c). (c) Maxwell-Lindblad simulation of the propagation pulse with $t_{\text{FWHM}}$~=~4.2~\unit{\mu s} and $\Gamma_{\text{hole}}$~=~300~\unit{kHz} at various $\alpha L$, which shows the distortion caused by the amplified frequency regions in (b). The experimental conditions in FIG~\ref{fig:fastlight_result} correspond to $\alpha L = -1.6$}
    \label{fig:disto_cont}
\end{figure*}

The resulting time advancement compared to propagation in vacuum is:
\begin{equation}\label{eq:adv}
    \Delta t =  \frac{\alpha L}{\pi^2}\left(\frac{1}{\Gamma_{\text{hole}}}- \frac{1}{\Gamma_{g}}\right)
\end{equation}
It is then evident that the stronger the gain and the narrower the spectral hole between the gain regions, the stronger the effect on the dispersion and thus the group velocity. 

We will now briefly describe, first, how the gain structure is prepared, second, how  amplified spontaneous emission in the created structures may limit the achievable dispersion and thereby also the time advancement of the pulse and, finally, aspects on pulse distortion. Additional details of these aspects are also provided in the supplementary material. 
The rare-earth ions are inhomogeneously broadened, such that different ions absorb at different frequencies. Further, the ions have three different and very long-lived (hours/days) doubly degenerate ground hyperfine levels at zero magnetic field. Thus, it is possible to use optical pumping to spectrally tailor the absorption profile until the structure seen in the right inset of FIG 1 is created, except that the peaks have not yet been inverted and are therefore absorption peaks. In principle, ions far detuned from the absorption peaks also contribute to the refractive index and has been neglected in Eq~(\ref{eq:nu}), but as shown in the supplementary, the effect of these can be neglected. The laser used has a linewidth of a few~\unit{kHz}  and the homogeneous broadening of the ions in the crystal is on the order of \unit{kHz}, hence the created spectral structure is fairly well represented by step functions, as seen in FIG~\ref{fig:fastlight_result}. 
The absorption peaks consist of nine different ion classes, where each ion class refers to one of the nine transitions between the three ground states and three upper-states hyperfine levels. Since these transitions function as an effective two-level system for the inversion, they must be coherently excited using a single pulse. Furthermore, due to limited available laser power, the peaks must consist entirely of transitions with strong oscillator strength such that complete inversion can be achieved coherently well within the lifetime of the excited state. An optical transparency is first created by removing all ions in a frequency range, subsequently ions belonging to suitable ion classes are pumped back and collected by targeting specific frequency ranges. The excited state lifetime in the Eu-system is long, 2~\unit{ms}. However, in a long strongly inverted medium, amplified spontaneous emission (ASE) will reduce the upper state lifetime. Spontaneous emission events at the beginning of the inverted region will be amplified as they travel along the inverted region and stimulate emission from the upper state. This effect scales exponentially with the magnitude of the gain. Hence there is a maximal allowable $\alpha L_{\text{max}}$, which can be estimated considering that the spontaneous emission that occurs at the very first slice of the crystal will have the dominant effect in reducing the lifetime. The details are given in the supplement and the result is,
\begin{equation}
    \alpha L_{\text{max}} = \ln\left(\frac{\gamma_{\text{rad,21}}}{W_{12}}\frac{z_{\text{R}}\lambda}{4n\pi L^2}\right)
\end{equation}
Where $\gamma_{\text{rad}, 21}$ is the radiative decay on the $^5D_0 \rightarrow~^7F_0$ transition used in the experiment, $z_\text{R}$, the Rayleigh length, $L$, the crystal length and $W_{12}$, the stimulated emission rate. In our case $z_{\text{R}} \approx L$ was required to ensure full inversion, and from simulations it followed that decreasing the lifetime by a factor of 2 was acceptable as this would leave enough time for first the inversion pulse and then the propagation pulse. This in essence gives a fundamental limit for the maximal gain and hence also the achievable dispersion in the material. In Eu:YSO, this limit is $\alpha L_{\text{max}} \approx 18$. In our experiments, we were, however, limited by the available Rabi frequency and thus could only invert gain structures with an $\alpha L_{\text{max}} \approx 2$. Thus, with an improved experimental setup it is possible to increase the time-bandwidth product even further.

Let us now, finally, turn to the topic of distortion. To have a distortion free pulse the bandwidth of a pulse, $\nu_{\text{FWHM}}$, must be smaller than the transmission window bandwidth, $\Gamma_{\text{hole}}$, and must remain below $\Gamma_{\text{hole}}$ by a factor that depends on the gain $\alpha L$.  The reason for this is that a pulse with a Gaussian envelope in time, will also have a Gaussian envelope in frequency. It will have frequency content in the wings which extends into the inverted region. This means that part of the Gaussian pulse, however small, will get selectively amplified, which may distort the pulse. However, the Gaussian envelops falls off fast enough that beyond a certain bandwidth, the gain structures causes practically no distortion, i.e., there is a level of tolerance when it comes to power leaking into the structure, before the pulse gets noticeably distorted, see FIG~\ref{fig:disto_cont}. This tolerance is connected to the magnitude of the gain of the structure, the frequency width of the optical window and the frequency width of the pulse. By multiplying the $\vec{E}$-field of the Gaussian in frequency space with the gain and comparing the area under the pulse to that of the unperturbed pulse, one can get a quick and intuitive picture whether the distortion will be significant. The error depends on the magnitude of the gain and the product of the $t_{\text{FWHM}}$ and hole-width $\Gamma_{\text{hole}}$: 
\begin{equation}\label{eq:erf}
    \xi = (1-e^{-\alpha L})\left(\erf\left(\frac{\pi t_{\text{FWHM}}\cdot\Gamma_{\text{hole}}}{4\sqrt{\text{ln}(2)}}\right)-1\right)
\end{equation}\newline
$\xi$ is then used as a measure of how much the pulse is amplified or distorted, with the error function, $\erf ()$ . Consequently there is a gain versus bandwidth trade-off, the largest time-bandwidth product is achieved at the highest $\alpha L$, but this structure will require higher Rabi frequency of the inversion pulse and reduce the tolerance for power in the structure. This tolerance can be seen in FIG \ref{fig:disto_cont} (a), where the increase in distortion due to the local amplification of the pulses', is plotted as a function of $\alpha L$ and the product of $\Gamma_{h}$ and $t_{\text{FWHM}}$. In time, this results initially in amplification and at larger values a ringing trailing the pulse front, which we both refer to as distortion for simplicity.

To conclude, we first note that semi-permanent frequency-domain tailoring of absorption structures rare earth ion doped crystals have been extensively used in quantum computing~\cite{Rippe2005}, quantum memories~\cite{NILSSON2005, Afzelius2009, Safavi2010}, spectral filtering~\cite{Beavan2013, Kinos2016}. More recently it has also been used for dispersion control to, e.g. enable dynamic speed of light control \cite{Li2017} or dynamic frequency shift control \cite{Li2016}. In this work it has been used for negative group velocity control, where a relative pulse advancement of $\frac{\Delta t}{t_{\text{FWHM}}}$ = 13$\%$ was obtained. This result surpasses previous negative group velocity work in terms of combined advancement and low pulse distortion. Further, simulations indicate that $\sim$ 30$\%$ advancement should be achievable with this system and other approaches, e.g., a concatenated sequence of spectrally tailored crystals, may enable still more significant pulse advances. This work also supports the more general statement that rare earth doped crystals constitute excellent test beds for exploiting the possibilities of using strongly dispersive materials for light propagation applications.

\bibliography{bib}

\section*{Supplement}
\subsection{Experimental setup}
The experiments were conducted in a Y$_2$SiO$_5$ crystal with dimensions 14$\times$15$\times$21~\unit{mm} doped to 1~at$\%$ natural abundance Europium containing both isotopes 151 and 153 in essentially equal proportion. The experiment was performed in the temperature range 2-4~\unit{K} in a custom built closed cycle cryostat made by MyCryofirm.

\begin{figure}[H]
    \centering
    \includegraphics[width = 0.45\textwidth]{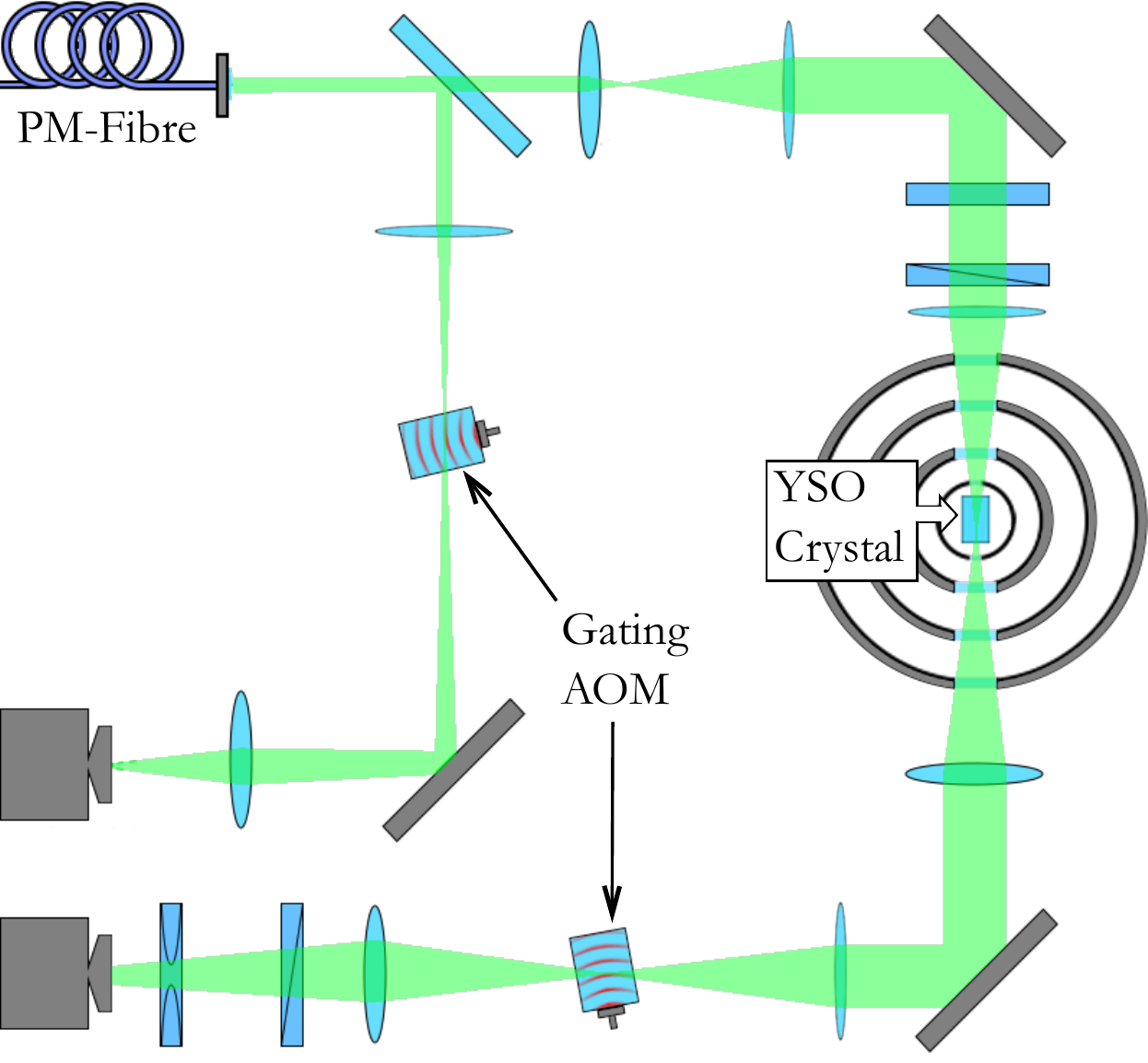}
    \caption{Optical setup during the experiment. }
    \label{fig:setupp}
\end{figure}

The light pulses in this work were measured with a Hamamatsu S5973-02 photodiode and amplified by a femto - DHPCA-100  trans-impedance amplifier. Due to the low light power in the propagation pulses it was necessary to use high gain on the transimpedence amplifier. This limits the bandwidth of the detection to $f_{\text{BW}}= 1.8~\unit{MHz}$ and rise time to, $t_{\text{rise}={0.35}/{f_{\text{BW}}}}=0.2~\unit{\mu s}$, which is significantly faster than the rise time of the pulse, which for a Gaussian is $t_{\text{rise, Gaussian}}=0.716~t_{\text{FWHM}}=3.01~\unit{MHz}$. From simulations we also know that this is sufficient bandwidth to detect any of the distortions could that arise from clipping the structure, should they occur (see FIG~\ref{fig:disto_cont}.c).
The optical setup is schematically shown in FIG~\ref{fig:setupp}, the light path is as follows, the Gaussian pulse come out of the PM-fibre onto a beamsplitter that splits it into the reference path, where it is focused on an AOM that gates the propagation pulse (but blocks the inversion pulse) and then finally it is focused on the detector. The rest of the light goes through the crystal. It is first expanded, the polarisation is aligned with the appropriate crystal axis and filtered and then focused down into the crystal in the cryostat, which is at 2~K. After exiting the cryostat it is collimated and then gated by an AOM before it is focused on the detector through a polariser and a filter that removes the fluorescence of the crystal and otherwise scattered light (e.g. from reflections of optics and cryostat windows). Significant care was put into adjusting the AOM delay such that pulses arrived at the detectors at the same time when no structure was prepared. 

\subsection{Preparation of structure}

\begin{figure*}
    \centering
    \includegraphics[width = 0.98\textwidth]{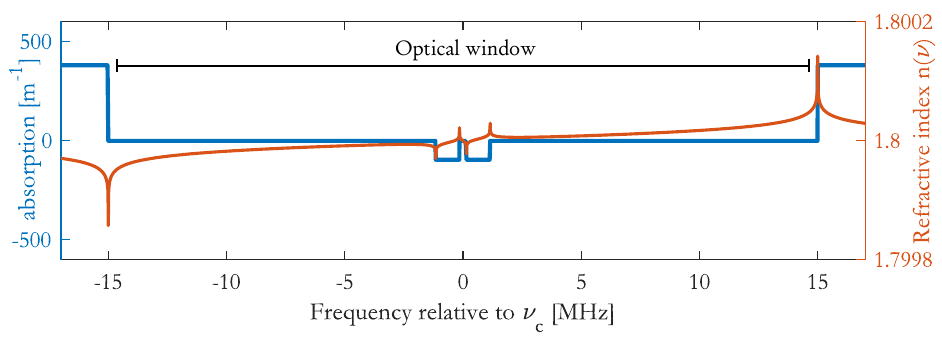}
    \caption{Simulated Dispersion effect of the gain structure and optical window}
    \label{fig:widepit}
\end{figure*}

Experiments were performed on the $^7F_0\rightarrow $~$^5D_0$ transition. A six level simulator was used to keep track of the ions and determine the optimal frequencies for the optical pumping and for creating the absorption and transmission structures, such that the ions that make up the structure have their strong transition at these frequencies. Ion classes with a weak transition in these frequency ranges are not targeted for the burn back. A 29~\unit{MHz} wide transmission widow was first created using optical pumping.  Two absorbing structures each 1~\unit{MHz} wide, centred at $\pm$ 650~\unit{kHz} from the centre of the optical transmission window, leaving a 300~\unit{kHz} transmission window at the centre frequency were created see FIG~\ref{fig:widepit}. The collected ions correspond to the transitions $\ket{1/2g}\rightarrow \ket{5/2e}$, $\ket{3/2g}\rightarrow \ket{3/2e}$ and $\ket{5/2g}\rightarrow \ket{1/2e}$, which have the strongest oscillator strengths and hence will be easier to invert~\cite{Cruzeiro2018}. If the ions with low oscillator strength would not be removed, these ions would absorb part of the energy from the inversion pulse, but would not reach inversion. This means that there would be an absorbing component in the structure causing the effective gain to be lower and as such the advancement would be reduced. The population inversion was then simulated using a Maxwell-Lindblad simulator, which used a simplified two level system where the oscillator strength were averaged corresponding to the fraction of the respective ions collected and using the estimated available laser power.  The two gain regions are inverted using a "sech-scan pulse", a frequency scan spliced in the middle of a sechhyp pulse \cite{Park2009}. A Gaussian pulse, $t_{\text{FWHM}}$~=~4.2~\unit{\mu s}, was used as the propagation pulse. From the Maxwell-Lindblad simulations the parameters for the experiment were determined, as well as how the distortion would manifest itself, due to the pulse power in the gain structures exceeding the distortion free limit. This can be seen in FIG~\ref{fig:disto_cont}, where various magnitudes of $\alpha L$ are shown. 

\subsection{Distortion}
To get a quick and intuitive idea of what will happen to the pulse for certain $t_{\text{FWHM}}$ and $\Gamma_{\text{hole}}$ with a given $\alpha L$ one can look at the pulse area in frequency space, after propagation through the gain structure. For this intuitive picture we do not consider the phase shift. This will give a clear representation of which frequency content is amplified and how large it is compared to the original envelope, see FIG \ref{fig:disto_cont}. To derive a quantitative expression for the distortion, we start with the ${E}$-field of a Gaussian pulse in time, 
\begin{equation}
    E(t) = exp\left({-\frac{4\text{ln}(2)t^2}{t_{\text{FWHM}}^{2}}}\right)
\end{equation}
Which naturally remains Gaussian in frequency,
\begin{equation}
    E(\nu) = \frac{t_{\text{FWHM}}\sqrt{\pi}}{2\sqrt{ln(2)}}exp\left({-\frac{t_{\text{FWHM}}^2\pi^2\nu^2}{4ln(2)}}\right)
\end{equation}
One can then write the expression for the gain structure as a function of frequency $\gamma (\nu)$. Since a Gaussian quickly drops off in amplitude, the effect of the gain structure will mostly come from the frequencies closest the pulse centre frequency, and the frequency content far in the wings will have little influence on the distortion. Thus the product of the amplification and the power in the wing of the pulse will be negligible already before the end of the proposed frequency width of the gain structure. This allows us to simplify the gain structure to be infinitely broad but with the transmission window around the centre frequency. Hence $\gamma(\nu)$:

\begin{equation}
    \gamma(\nu) = exp\left(-\alpha L \left(\theta\left(-\frac{\Gamma_h}{2}+\nu\right)+\theta\left(-\frac{\Gamma_h}{2}-\nu\right)\right)\right)
\end{equation}

Simply multiplying these and then integrating the product gives a measure of the energy in the pulse, then subtracting the original area leads to the amplified content:

\begin{equation}
    \xi = \frac{{\int_{-\infty}^{\infty}E(\nu)\gamma(\nu)-E(\nu)d\nu}}{\int_{-\infty}^{\infty}E(\nu)d\nu}
\end{equation}
Solving this lead to:
\begin{equation*}                                                                                                                      
    \xi = (1-e^{-\alpha L})\left(\erf\left(\frac{\pi t_{\text{FWHM}}\cdot\Gamma_{\text{hole}}}{4\sqrt{\text{ln}(2)}}\right)-1\right)
\end{equation*}

\subsection{Derivation of $n(\nu)$}
The refractive index for the transmission window and the material combined is:
\begin{equation}\label{eq:N}
    n(\nu) = n_{\text{host}}+\frac{\chi_{\text{real}}}{2n_{\text{host}}}
\end{equation} \cite[p.~377]{Siegman}

We start from the susceptibility:

\begin{equation}
\chi_h(\nu) =  -\frac{2j \alpha_0 c n_{\text{host}}}{\nu_{\text{central}}\pi^2(\Gamma_h+2j(\nu-\nu_0))}
\end{equation}

Assume that the absorption profile looks flat from the perspective of the pit/structure and that there is no background absorption in the pit:
\begin{multline}
        g(\nu) = \theta\left(-\frac{\Gamma_{\text{hole}}}{2}+\nu_0\right)+\theta\left(-\frac{\Gamma_{\text{hole}}}{2}-\nu_0\right)\\-\theta \left(-{\Gamma_{g}}-\nu_0 \right)-\theta\left(-\Gamma_{g}+\nu_0\right)
\end{multline}
Here $\theta(\nu)$ represents the Heaviside step-function. The real part of $\chi_h(\nu)$ describes the refractive index, the imaginary part the absorption. Then to get the $\Re(\chi_h(\nu))$ for the structure the integral of the susceptibility multiplied with the absorption profile must be computed:
\begin{equation}
    \chi_{\text{real}} = \int^{\infty}_{\infty} g(\nu)\Re(\chi_h(\nu))d\nu_0
\end{equation}
Furthermore assume that the structure terminates at $\Gamma_{g}$ and that beyond this point the absorption is far enough away to not have a significant influence. One could consider these effects by simply adding similar terms with the appropriate boundary conditions, i.e. an optical window with width 29~\unit{MHz}, but as can be seen from FIG~\ref{fig:fastlight_result}, right inset, and FIG~\ref{fig:widepit} it has little effect on the result. The simplified situation is described by:
\begin{equation}
    \chi_{\text{real}} = \int^{-{\Gamma_{\text{hole}}}/{2}}_{-\Gamma_{g}}\Re(\chi_h(\nu))d\nu_0+\int^{\Gamma_{g}}_{{\Gamma_{\text{hole}}}/{2}}\Re(\chi_h(\nu))d\nu_0
\end{equation}
Then computing this integral this gives:
\begin{equation}
\begin{split}
    \chi_{\text{real}} = -\frac{\alpha_0 c n_{\text{host}}}{2 \pi^2 \nu_{\text{central}}}  \\ \text{ln}\left(\frac{4\Gamma_{g}^2+8\Gamma_{g}\nu+4\nu^2+\Gamma_h^2}{4\nu^2+4\Gamma_{\text{hole}}\nu+\Gamma_h^2+\Gamma_{\text{hole}}^2}\right)  \\ +\text{ln}\left(\frac{4\nu^2-4\Gamma_{\text{hole}}\nu+\Gamma_h^2+\Gamma_{\text{hole}}^2}{4\Gamma_{g}^2-8\Gamma_{g}\nu+4\nu^2+\Gamma_h^2}\right)
\end{split}
\end{equation}
The homogeneous linewidth $\Gamma_h$ (=250$\pm$50~\unit{Hz}~\cite{Konz2003}), will be negligible compared to the other frequency widths so can be ignored giving:
\begin{equation}
\begin{split}
    \chi_{\text{real}} = -\frac{\alpha_0 c n_{\text{host}}}{2 \pi^2 \nu_{\text{central}}}\times\\ \text{ln}\left(\left(\frac{4\Gamma_{g}^2+8\Gamma_{g}\nu+4\nu^2}{4\nu^2+4\Gamma_{\text{hole}}\nu+\Gamma_{\text{hole}}^2}\right)\left(\frac{4\nu^2-4\Gamma_{\text{hole}}\nu+\Gamma_{\text{hole}}^2}{4\Gamma_{g}^2-8\Gamma_{g}\nu+4\nu^2}\right)\right)
    \end{split}
\end{equation}
using equation~(\ref{eq:N}) and applying a Taylor expansion in $\nu$ around the centre of the optical window:
 \begin{equation}
n(\nu) = n_{\text{host}}+\frac{\alpha_0 c}{ \pi^2\nu_{\text{central}}}\left(\frac{1}{\Gamma_{g}}-\frac{2}{\Gamma_{\text{hole}}}\right)\nu
\end{equation}

\subsection{Un-normalised pulses and reference}
\begin{figure*}
    \centering
    \includegraphics[width = 0.98\textwidth]{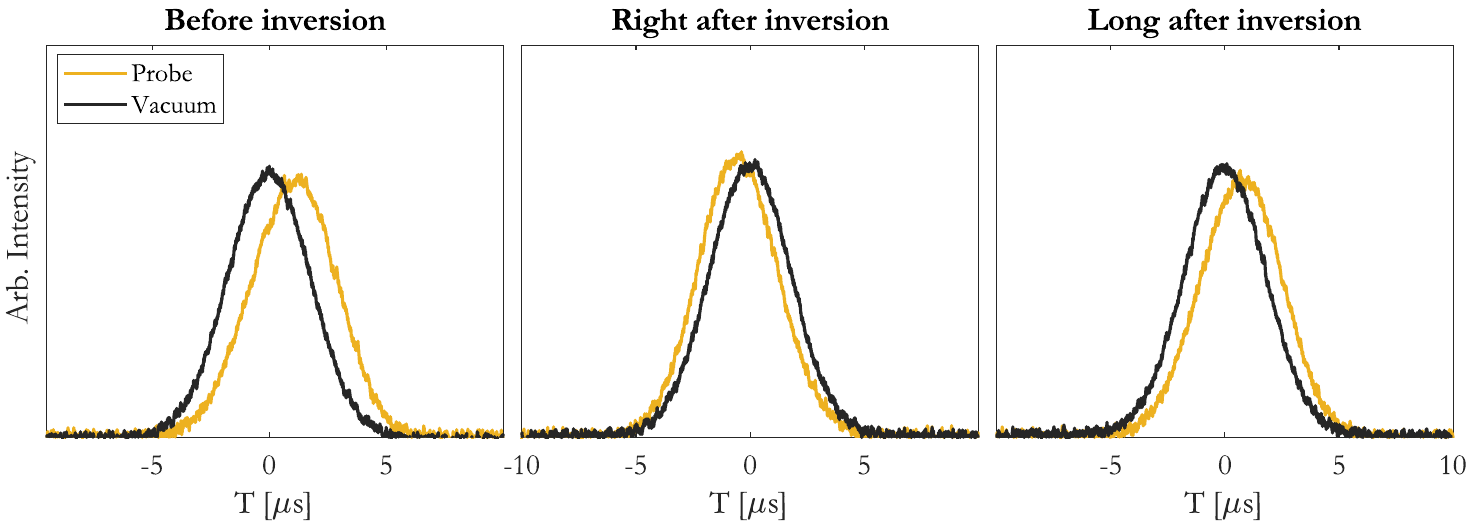}
    \caption{Probe pulse sequence, where the vacuum-data is scaled according to best effort using simulations as if the intensities in both optical arms were the same. Here vacuum refers to the pulse that does not travel through the crystal but is picked off as shown in FIG \ref{fig:setupp}}
    \label{fig:TrippleGauus}
\end{figure*}
In FIG~\ref{fig:TrippleGauus} the pulse sequence from the experiment is shown. After the structure is prepared, we sent in an initial pulse without inversion, to probe the collected $\alpha L$ without disturbing the structure. Between probe pulses a 15~\unit{ms} wait was added to allow proper relaxation. After which an inversion pulse is sent immediately followed by a probe pulse. Finally after another 15~\unit{ms} wait, a third probe pulse is sent to assess the structure after absorption. This data is scaled according to best efforts to have equal power in both optical paths. As one can see from the vacuum pulse compared to the advanced pulse there is a slight amplification of the advanced pulse, and minor attenuation of the slowed down pulse. In the main part of the paper the reference pulses are normalised to the probe pulses to more clearly display the temporal shift. 

\subsection{Maximal gain}
For these calculations we simplify the geometry, and assume the excited volume of the crystal to be roughly cylindrical. Then a spectral packet that originated at the very start of the crystal will reach the end with intensity:
\begin{equation}\label{eq:I}
    I_{ij} \approx \frac{\gamma_{\text{rad,ij}}h\nu_{ij} r^2}{4\sigma_{ij} L^2}e^{-\alpha_{ij} L}
\end{equation} \cite[p.~522]{Siegman}, where the subscripts indicate the specific levels involved. With beam radius, $r$, cross section of transition $i\rightarrow j$, $\sigma_{ij}$ and crystal length, $L$. The total radiative decay rate from $^5D_0$ is $\gamma_{\text{rad,tot}} = \frac{1}{2\text{ms}}=500$~\unit{Hz}. The radiative decay for a specific transition is $\gamma_{\text{rad,ij}}$, $r$ the radius of the inverted cylinder, the absorption on a particular transition, $\alpha_{ij}$. Note that $\alpha_{ij} <0$ when the population is inverted.
Then using the relation of the stimulated transition rate and intensity one can find the relationship between the radiative decay rate and the transition rate. The stimulated emission rate $W_{ij}$ between a level $i$ and $j$ is given by:
\begin{equation}
    W_{ij} = \frac{\sigma_{ij}I_{ij}}{h\nu}
\end{equation} combining these two equations leads to the ratio of the stimulated and radiative rates,
\begin{equation}
    \frac{W_{ij}}{\gamma_{\text{rad},ij}}\approx \left(\frac{r}{2L}\right)^{2}e^{-\alpha_{ij} L}
\end{equation}
 \cite[p.~522]{Siegman}.
The $^5D_0 \rightarrow~^7F_{0}$, transition is used for this experiment, but there is also relaxation to the other $^7F_J$ levels, hence it is important to know what dominates the stimulated emission. 
\begin{figure}[H]
    \centering
    \includegraphics[width = .45\columnwidth]{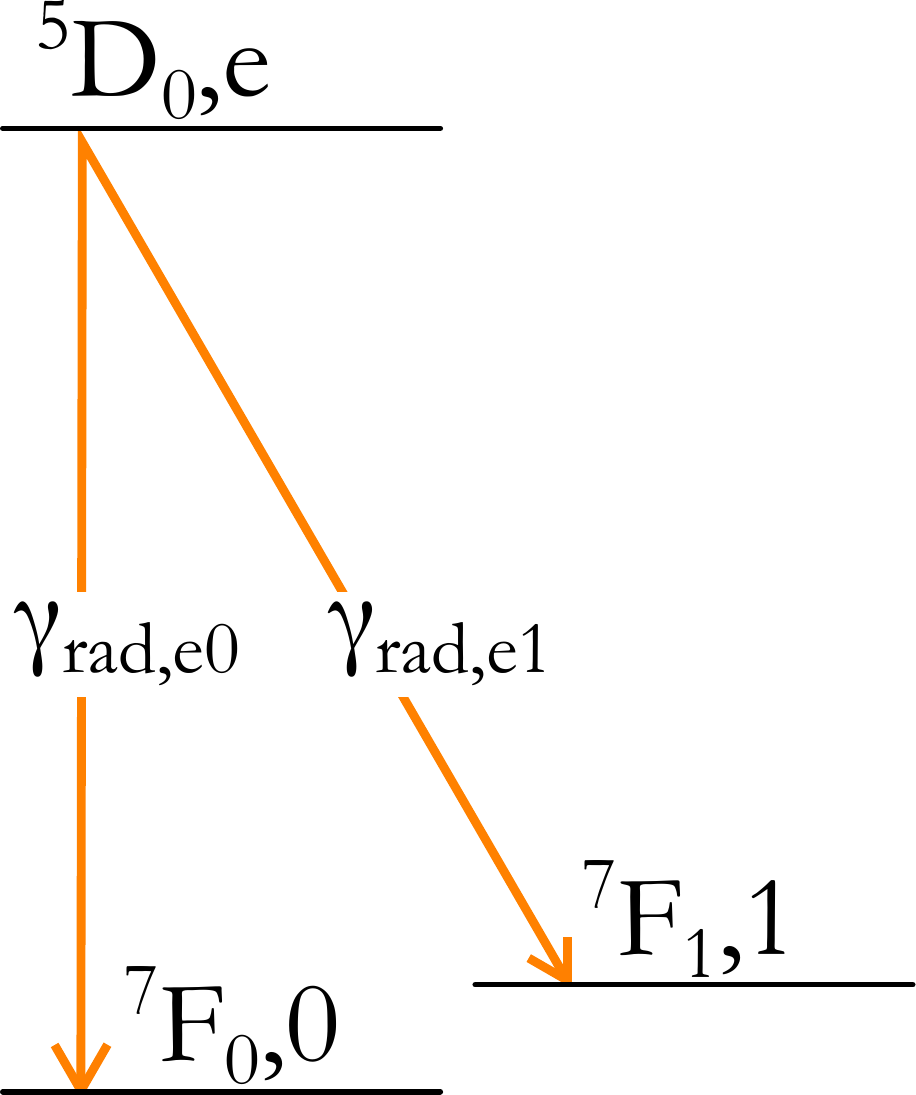}
    \caption{Relevant levels for radiative decay. The $^5D_0$ is denoted by $e$, the $^7F_0$ by $0$ and the $^7F_1$ by $1$}
    \label{fig:StarkLevels}
\end{figure} 
From the radiative decay rate for this transition, the radiative decay rate for the $^5D_0\rightarrow~^7F_0$, $\gamma_{\text{rad, ceo}}$ transition can be calculated from its oscillator strength $F_{e0} = 1.3\cdot 10^{-8}$~\cite{Graf1998}, as 
\begin{equation}
    \gamma_{\text{rad,e0}}= F_{e0} 3^*\gamma_{\text{rad, ceo}} 
\end{equation} \cite[p.~122]{Siegman}, \cite{Konz2003} $\gamma_{\text{rad,ceo}}$ is the radiative decay rate for the classical electron oscillator, \cite[p.~122]{Siegman}
\begin{equation}
    \gamma_{\text{rad,ceo}} = 2.47\cdot10^{-22}n\cdot\nu_{ij}^{2}
\end{equation}
with the refractive index $n = 1.8$, for $Y_2SiO_5$~\cite{Lauritzen2012}. The dimensionless polarisation overlap factor $3^*$ for site 1 in Eu:$^{3+}$Y$_2$SiO$_5$ with the light polarisation aligned with the D1 axis is given by:
\begin{equation*}
3^*=\frac{3.9}{\sqrt{3.9^2+.8^2}}*3 = 2.94    
\end{equation*}
The radiative decay rate can now be calculated, $\gamma_{\text{rad,e0}}=4.23$ and the branching ratio for the $^5D_0\rightarrow~^7F_0$ transition is $\gamma_{\text{rad,e0}}/\gamma_{\text{rad,tot}}=0.8\%$, which means that 99.2$\%$ from the $^5D_0$ relaxes through the $^7F_{J\in 1,2,3,4,5,6}$ lines. The inverted structure on the  $^5D_0\rightarrow~^7F_0$ transition is 2.3~\unit{MHz} wide and since the  $^5D_0\rightarrow~^7F_{j\in 1,2,3,4,5,6}$ are much broader, the cross-section, $\sigma$ and the emission coefficient, $\alpha_{ij} = \Delta N_{ij}\sigma_{ij}$ for the stimulated emission will be scaled down with the ratio of the linewidth. The strongest emission from these transitions is on the  $^5D_0\rightarrow~^7F_1$ see figure 3 in \cite{Konz2003} It is also considerably narrower then the other transitions. Könz et al. report it to be much narrower than 6~\unit{GHz} \cite{Konz2003}. The cross-section for stimulated emission for the  $^5D_0\rightarrow~^7F_1$, is $\sigma_{e0}=1.2\cdot 10^{-18}$~\unit{m^2} \cite{Krivachy2023}. We will calculate what linewidth the  $^5D_0\rightarrow~^7F_1$ would need to have in order to have the same cross-section as the  $^5D_0\rightarrow~^7F_0$ transition. The cross-section $\sigma_{ij}$ is given by p288 in \cite{Siegman}.
\begin{equation}\label{eq:lorri}
    \sigma_{ij} = \frac{3^*\gamma_{\text{rad,ij}}}{\Delta\nu_{\text{Lorentzian}}}\left(\frac{\lambda_{ij}}{n}\right)^2
\end{equation}
If $\sigma_{e1}$ where to be equal to $\sigma_{e0}$ this would require a linewidth of,
\begin{equation}
    \Delta\nu_{\text{Lorentzian}} = \frac{3^*}{\sigma_{e0}}\left(\frac{\lambda_{e1}}{n}\right)^2\gamma_{\text{rad,e1}}
\end{equation} The radiative decay rate for the  $^5D_0\rightarrow~^7F_1$ was estimated to be $\gamma_{\text{rad,el}}\approx 40$ from the relative line areas in figure 1b in \cite{Yano1991}, due to the limited resolution this was a rather crude estimate. The wavelength of the $^5D_0\rightarrow~^7F_1$ transition is $\lambda_{e1} = 586.882$~\unit{nm} \cite{Konz2003}. The required $^5D_0\rightarrow~^7F_1$ linewidth was then calculated from \ref{eq:lorri} using  $3^*=3$ as an upper bound which gives $\Delta\nu_{\text{Lorentzian}} = 10~\unit{MHz}$ Since it is very unlikely that the transition is this narrow one can assume that the stimulated ASE decay is completely dominated by the $^5D_0\rightarrow~^7F_0$. So we can simply take
\begin{equation}
    \frac{W_{e0}}{\gamma_{\text{rad,e0}}} \approx \left(\frac{\sqrt{\frac{z_{\text{R}}\lambda_{e0}}{n\pi}}}{2L}\right)^2\cdot e^{-\alpha_{e0} L}
\end{equation}
Where $z_{\text{R}}$ is the Raleigh range $\lambda_{e0}$ the wavelength, $W_{e0}$ the stimulated emission rate. Hence, given an acceptable reduction of the lifetime, given by the ratio $W_{e0}/\gamma_{\text{rad,e0}}$, one can determine the maximum $\alpha L_{\text{max}}$:
\begin{equation}
    \alpha L_{\text{max}} = \ln\left(\frac{\gamma_{\text{rad,e0}}}{W_{e0}}\frac{z_{\text{R}}\lambda}{4n\pi L^2}\right)
\end{equation}
If we allow for a reduction of the upper lifetime to half, i.e. $\frac{w_{e0}}{\gamma_{\text{rad,e0}}}=1$ we get $\alpha L_{\text{max}} \approx 18 $ for our case. From either FIG~\ref{fig:disto_cont}~A or by plugging it in equations~(\ref{eq:adv}) and (\ref{eq:erf}) given the error tolerance one can then see an advancement of $\approx 30\%$ can be obtained.

\end{document}